\documentclass[amsmath,aps,showpacs,a4paper,10pt]{revtex4}

 \usepackage{epsf}
 \usepackage{graphicx}    

\begin{document}

 \newcommand{\be}[1]{\begin{equation}\label{#1}}
 \newcommand{\ee}{\end{equation}}
 \newcommand{\bea}{\begin{eqnarray}}
 \newcommand{\eea}{\end{eqnarray}}
 \def\disp{\displaystyle}

 \def\gsim{ \lower .75ex \hbox{$\sim$} \llap{\raise .27ex \hbox{$>$}} }
 \def\lsim{ \lower .75ex \hbox{$\sim$} \llap{\raise .27ex \hbox{$<$}} }

 \begin{titlepage}

 \begin{flushright}
 arXiv:0907.2749
 \end{flushright}

 \title{\Large \bf Varying Alpha Driven by
 the Dirac-Born-Infeld Scalar Field}

 \author{Hao~Wei\,}
 \email[\,email address:\ ]{haowei@bit.edu.cn}
 \affiliation{Department of Physics, Beijing Institute
 of Technology, Beijing 100081, China}

 \begin{abstract}\vspace{1cm}
 \centerline{\bf ABSTRACT}\vspace{2mm}
 Since about ten years ago, varying $\alpha$ theories attracted
 many attentions, mainly due to the first observational
 evidence from the quasar absorption spectra that the fine
 structure ``constant'' might change with cosmological
 time. In this work, we investigate the cosmic evolution of
 $\alpha$ driven by the Dirac-Born-Infeld (DBI) scalar field.
 To be general, we consider various couplings between the DBI
 scalar field and the electromagnetic field. We also confront
 the resulting $\Delta\alpha/\alpha$ with the observational
 constraints, and find that various cosmological evolution
 histories of $\Delta\alpha/\alpha$ are allowed. Comparing
 with the case of varying $\alpha$ driven by quintessence, the
 corresponding constraints on the parameters of coupling have
 been relaxed, thanks to the relativistic correction of the
 DBI scalar field.
 \end{abstract}


 \maketitle

 \end{titlepage}

 \renewcommand{\baselinestretch}{1.4}


\section{Introduction}\label{sec1}

For many years, there are some unremitting speculations in the
 subject of the possible variations of fundamental constants.
 One of the earliest works is the famous large number
 hypothesis proposed by Dirac in 1937~\cite{r1}. In the
 fundamental ``constants'', the most observationally sensitive
 one is the electromagnetic fine structure ``constant'',
 $\alpha=e^2/\hbar c$. Since about ten years ago, this subject
 attracted many attentions again, mainly due to the first
 observational evidence from the quasar absorption spectra that
 the fine structure ``constant'' might change with cosmological
 time~\cite{r2,r3}.

Subsequently, many authors obtained various observational
 constraints on the possible variation of the fine structure
 ``constant'' $\alpha$. In the literature, it is convenient to
 introduce a quantity
 $\Delta\alpha/\alpha\equiv (\alpha-\alpha_0)/\alpha_0$,
 where the subscript ``0'' indicates the present value of the
 corresponding quantity. Obviously, $\Delta\alpha/\alpha$ is
 time-dependent. A brief summary of the observational
 constraints on $\Delta\alpha/\alpha$ can be found in
 e.g.~\cite{r4}. The most ancient constraint comes from the
 Big Bang Nucleosynthesis (BBN)~\cite{r5,r6}, namely,
 $|\Delta\alpha/\alpha|\,\lsim\, 10^{-2}$, in the redshift
 range $z=10^{10}-10^8$. The next constraint comes from the
 power spectrum of anisotropy in the cosmic microwave
 background (CMB)~\cite{r6}, i.e.,
 $|\Delta\alpha/\alpha| < 10^{-2}$, for redshift
 $z\simeq 10^3$. In the medium redshift range, the constraint
 comes from the absorption spectra of distant
 quasars~\cite{r2,r3,r7,r8}. Since the results in the
 literature are controversial, it is better to consider the
 conservative constraint
 $|\Delta\alpha/\alpha|\,\lsim\, 10^{-6}$~\cite{r4}, in the
 redshift range $z=3-0.4$. From the radioactive life-time of
 $^{187}$Re derived from meteoritic studies~\cite{r9}, the
 constraint is given by $|\Delta\alpha/\alpha|\,\lsim\, 10^{-7}$
 for redshift $z=0.45$. Finally, from the Oklo natural nuclear
 reactor~\cite{r10}, it is found that
 $|\Delta\alpha/\alpha|\,\lsim\, 10^{-7}$ for redshift $z=0.14$.
 For convenience, we summarize the above constraints in
 Table~\ref{tab1} and label them by the gray areas in
 Figs.~\ref{fig2}---\ref{fig5}, \ref{fig7} and \ref{fig8}.

 \begin{table}[htbp]
 \begin{center}
 \vspace{3mm}  
 \begin{tabular}{cccc} \hline\hline
 ~~$|\Delta\alpha/\alpha|$~~~ & ~~~redshift~~~
 & ~~~observation~~~ & \hspace{6mm} Ref.\hspace{6mm} \\ \hline
 $\lsim\, 10^{-2}$ & $10^{10}-10^8$ & BBN & \cite{r5,r6}\\
 $< 10^{-2}$ & $10^3$ & CMB & \cite{r6}\\
 $\lsim\, 10^{-6}$ & $3-0.4$ & quasars & \cite{r2,r3,r7,r8}\\
 $\lsim\, 10^{-7}$ & $0.45$ & meteorite & \cite{r9}\\
 $\lsim\, 10^{-7}$ & $0.14$ & Oklo & \cite{r10}\\
 \hline\hline
 \end{tabular}
 \end{center}
 \caption{\label{tab1} The observational constraints on
 $\Delta\alpha/\alpha$.}
 \end{table}

A varying $\alpha$ might be due to a varying speed of light
 $c$~\cite{r11,r12,r13}, while Lorentz invariance is broken.
 The other possibility for a varying $\alpha$ is due to a
 varying electron charge $e$. In 1982, Bekenstein proposed
 such a varying $\alpha$ model~\cite{r14}, which preserves
 local gauge and Lorentz invariance, and is generally
 covariant. This model has been revived and generalized after
 the first observational evidence of varying $\alpha$ from the
 quasar absorption spectra~\cite{r2,r3}. This is a dilaton
 theory with coupling to the electromagnetic $F^2$ part of the
 Lagrangian, but not to the other gauge fields. One example of
 this type of models is the so-called BSBM model in the
 literature~\cite{r15,r16,r17,r18}.

On the other hand, dark energy~\cite{r19} has been one of the
 most active fields in modern cosmology since the discovery of
 accelerated expansion of our universe~\cite{r20}. Most of dark
 energy models are described by a dynamical scalar field. It is
 possible to image that such a cosmological scalar field could
 be coupled with the electromagnetic field, and hence could
 drive the variation of $\alpha$. So, one can generalize
 the Bekenstein-type varying $\alpha$ model by replacing the
 dilaton with the scalar field dark energy. Further, the
 coupling between the scalar field and the electromagnetic
 field could also be generalized. Actually, the varying
 $\alpha$ models driven by quintessence have been extensively
 investigated in the literature
 (e.g.~\cite{r4,r21,r22,r23,r24,r25,r26,r27,r36}). In addition,
 we mention that varying $\alpha$ driven by phantom has been
 considered in the BSBM model~\cite{r15,r16,r17,r18} while its
 model parameter $\omega$ is negative. The special case of
 varying $\alpha$ driven by k-essence whose Lagrangian
 ${\cal L}(X)=X^n-V(\phi)$ has also been considered in
 e.g.~\cite{r21}.

Recently, the Dirac-Born-Infeld (DBI) scalar field attracted
 many attentions. In type IIB string theory, the DBI action
 arises naturally in the D3-brane motion within a warped
 geometry or ``throat''. It can give a variety of novel
 cosmological consequences. For instance, the DBI scalar field
 can be used to drive the inflation (see
 e.g.~\cite{r28,r29,r30,r31}). More recently, the DBI scalar
 field has been proposed to play the role of dark
 energy~\cite{r32,r33,r34,r35}. Therefore, it is natural to
 consider the varying $\alpha$ driven by the DBI scalar field
 in the present work.

This paper is organized as followings. In Sec.~\ref{sec2}, we
 will briefly review the varying $\alpha$ driven by
 quintessence. In Sec.~\ref{sec3}, we consider the varying
 $\alpha$ driven by the DBI scalar field, and confront it with
 the observational constraints. A brief conclusion is given
 in Sec.~\ref{sec4}.


\section{Brief review on the varying alpha driven by
 quintessence}\label{sec2}

Following~\cite{r27,r4,r21}, the relevant action is given by
 \be{eq1}
 S=\frac{1}{2m_p^2}\int d^4x\,\sqrt{-g}\,R+
 \int d^4 x\,\sqrt{-g}\,{\cal L}_\phi-
 \frac{1}{4}\int d^4 x\,\sqrt{-g}\,B_F(\phi)\,F_{\mu\nu}F^{\mu\nu}+
 S_m+S_r\,,
 \ee
 where  $F_{\mu\nu}$ are the components of the electromagnetic
 field tensor; $S_m$ and $S_r$ are the actions of pressureless
 matter and radiation, respectively; $m_p\equiv(8\pi G)^{-1/2}$
 is the reduced Planck mass; ${\cal L}_\phi$ is the Lagrangian
 of the scalar field. For the case of quintessence, the
 corresponding ${\cal L}_\phi$ reads
 \be{eq2}
 {\cal L}_\phi=
 \frac{1}{2}\partial_\mu\phi\partial^\mu\phi-V(\phi)\,,
 \ee
 where $V(\phi)$ is the potential. Notice that $B_F$ takes the
 place of $e^{-2}$ in Eq.~(\ref{eq1}) actually~\cite{r22,r37},
 one can easily see that the effective fine structure
 ``constant'' is given by~\cite{r4,r21}
 \be{eq3}
 \alpha=\frac{\alpha_0}{B_F(\phi({\bf x},t))}\,.
 \ee
 Thus, we find that
 \be{eq4}
 \frac{\Delta\alpha}{\alpha}\equiv\frac{\alpha
 -\alpha_0}{\alpha_0}=\frac{1-B_F(\phi)}{B_F(\phi)}\,.
 \ee
 Notice that the present value of the coupling $B_F$ should be
 $1$. In general, $\phi$ and hence $\alpha$ are functions of
 spacetime. However, as is well known, we can safely neglect
 the spatial variation of $\phi$ and $\alpha$, which is
 usually a good approximation. Therefore, we only consider the
 homogeneous $\phi$ and $\alpha$ throughout this paper. The
 relevant equations governing the cosmological evolution in
 a flat universe read
 \be{eq5}
 H^2=\frac{1}{3m_p^2}\left(\rho_m+\rho_r+\rho_\phi\right),
 \ee
 \be{eq6}
 \ddot{\phi}+3H\dot{\phi}+V_{,\phi}=0\,,
 \ee
 where $H\equiv\dot{a}/a$ is the Hubble parameter;
 $a=(1+z)^{-1}$ is the scale factor (we have set $a_0=1$);
 $z$ is the redshift; a dot denotes the derivative with respect
 to cosmic time $t$; the energy densities of pressureless
 matter and radiation are given by $\rho_m=\rho_{m0}a^{-3}$ and
 $\rho_r=\rho_{r0}a^{-4}$, respectively; $\rho_\phi$ is the
 energy density of scalar field $\phi$ [for the case of
 quintessence, $\rho_\phi=\dot{\phi}^2/2+V(\phi)$]; the
 subscript ``$,\phi$'' denotes the derivative with respect
 to $\phi$. In fact, due to the coupling between the scalar
 field and the electromagnetic field, there should be an
 additional term in the right hand side of the equation of
 motion for $\phi$, namely Eq.~(\ref{eq6}). This additional
 term is proportional to $F_{\mu\nu}F^{\mu\nu}$ and the
 derivative of $B_F$~\cite{r21}. However, it can be safely
 neglected thanks to the following facts: (i) the derivative of
 $B_F$ is in fact equivalent to the time derivative of $\alpha$
 [cf. Eq.~(\ref{eq3})], which is very small (given equivalence
 principle constraints~\cite{r27}); see e.g.~\cite{r21};
 (ii) the statistical average of the term
 $F_{\mu\nu}F^{\mu\nu}$ over a current state of the universe
 is zero~\cite{r4}.

One can numerically solve Eqs.~(\ref{eq5}) and (\ref{eq6}) to
 obtain the cosmological evolution of $\phi(t)$. Then, the
 corresponding $\alpha(t)$ is ready. We can confront it with
 the observational constraints. In fact, the varying $\alpha$
 models driven by quintessence have been extensively
 investigated in the literature
 (e.g.~\cite{r4,r21,r22,r23,r24,r25,r26,r27}). In the next
 section, we turn to the case of DBI scalar field.


\section{Varying alpha driven by the DBI scalar field}\label{sec3}

In this section, we consider the varying $\alpha$ driven by the
 DBI scalar field. We firstly give out the relevant equations
 and solve them to get $\phi(t)$ and hence $\alpha(t)$. Then,
 we confront the fine structure ``constant'' $\alpha$ with the
 observational constraints.


\subsection{Equations}\label{sec3a}

The Lagrangian of DBI scalar field is given by~\cite{r28,r33,r34}
 \be{eq7}
 {\cal L}_\phi=-\frac{1}{g_{_{\rm YM}}^2}\left[
 f(\phi)^{-1}\sqrt{1+f(\phi)\,\partial_\mu\phi\partial^\mu\phi}
 -f(\phi)^{-1}+V(\phi)\right],
 \ee
 where $g_{_{\rm YM}}^2$ is the Yang-Mills coupling; $V(\phi)$
 is the potential; $T=f(\phi)^{-1}$ is the warped brane
 tension. The pressure and energy density of the DBI scalar
 field are given by~\cite{r28,r33,r34}
 \be{eq8}
 p_\phi=\frac{\gamma-1}{f\gamma}-V(\phi)\,,
 \ee
 \be{eq9}
 \rho_\phi=\frac{\gamma-1}{f}+V(\phi)\,,
 \ee
 where the Lorentz factor
 \be{eq10}
 \gamma=\frac{1}{\sqrt{1-f(\phi)\,\dot{\phi}^2}}\,,
 \ee
 which measures the ``relativistic'' motion of the DBI scalar
 field. In the ``non-relativistic'' limit, $K/T\ll 1$, and
 $\gamma\to 1+K/T$, where $K=\dot{\phi}^2/2$ is the canonical
 kinetic energy. In this case, the equation-of-state parameter
 (EoS) of DBI scalar field $w=p_\phi/\rho_\phi\to(K-V)/(K+V)$,
 i.e., the DBI scalar field reduces to an ordinary quintessence
 field~\cite{r34}. In the ``ultra-relativistic'' limit,
 $\gamma\to\infty$. In the medium $\gamma$ range, the
 non-canonical behavior due to the relativistic corrections
 will be crucial~\cite{r28,r33,r34}.

The equation of motion for the DBI scalar field
 reads~\cite{r28,r33}
 \be{eq11}
 \ddot{\phi}+\frac{3f_{,\phi}}{2f}\dot{\phi}^2-
 \frac{f_{,\phi}}{f^2}+\frac{3H}{\gamma^2}\dot{\phi}+
 \left(V_{,\phi}+\frac{f_{,\phi}}{f^2}\right)\frac{1}{\gamma^3}
 =0\,.
 \ee
 It is worth noting that the additional term in the right hand
 side of Eq.~(\ref{eq11}) due to the coupling between the
 scalar field and the electromagnetic field can be safely
 neglected, as already mentioned in Sec.~\ref{sec2}. The
 Friedmann equation is given by~\cite{r28,r33}
 \be{eq12}
 H^2=\frac{1}{g_{_{\rm YM}}^2}\cdot\frac{1}{3m_p^2}\left(\rho_m
 +\rho_r+\rho_\phi\right),
 \ee
 in which the corresponding $\rho_\phi$ is given in
 Eq.~(\ref{eq9}). For the AdS throat, $f(\phi)$ is given
 by~\cite{r28,r33,r34}
 \be{eq13}
 f(\phi)=\frac{\lambda}{\phi^4}\,,
 \ee
 where $\lambda$ is a dimensionless constant. As
 in~\cite{r33,r34}, here we consider a quadratic potential
 \be{eq14}
 V(\phi)=\frac{1}{2}m^2\phi^2\,,
 \ee
 where $m$ is a constant. It is convenient to introduce the
 following dimensionless quantities
 \be{eq15}
 \tilde{\phi}\equiv\frac{\phi}{m_p}\,,~~~~~
 \tilde{t}\equiv m t\,,~~~~~
 \tilde{m}\equiv\frac{1}{g_{_{\rm YM}}}\cdot\frac{H_0}{m}\,,~~~~~
 \tilde{\lambda}\equiv\frac{m^2}{m_p^2}\lambda\,.
 \ee
 Thus, we can recast Eq.~(\ref{eq12}) as
 \be{eq16}
 \tilde{H}^2=\left(\frac{a^\prime}{a}\right)^2=
 \tilde{m}^2\left(\Omega_{m0}\,a^{-3}+\Omega_{r0}\,a^{-4}+
 \Omega_{\phi 0}\frac{\tilde{\rho}_\phi}{\tilde{\rho}_{\phi 0}}
 \right),
 \ee
 where $\Omega_i\equiv\rho_i/(3m_p^2H_0^2)$ are the fractional
 energy densities of pressureless matter, radiation and DBI
 scalar field for $i=m$, $r$ and $\phi$, respectively; a prime
 denotes the derivative with respect to the new time variable
 $\tilde{t}$; and
 \be{eq17}
 \frac{\tilde{\rho}_\phi}{\tilde{\rho}_{\phi 0}}=
 \frac{\rho_\phi}{\rho_{\phi 0}}=\frac{\left(
 \tilde{\gamma}-1\right)\tilde{\phi}^4\tilde{\lambda}^{-1}
 +\tilde{\phi}^2 /2}{\left(\tilde{\gamma}_0-
 1\right)\tilde{\phi}_0^4\tilde{\lambda}^{-1}+
 \tilde{\phi}_0^2 /2}\,,
 \ee
 in which
 \be{eq18}
 \tilde{\gamma}=\gamma=\frac{1}{\sqrt{1-
 \tilde{\lambda}\tilde{\phi}^{-4}\tilde{\phi}^{\prime 2}}}\,.
 \ee
 Also, we recast Eq.~(\ref{eq11}) as
 \be{eq19}
 \tilde{\phi}^{\prime\prime}-
 \frac{6}{\tilde{\phi}}\tilde{\phi}^{\prime 2}+
 \frac{4}{\tilde{\lambda}}\tilde{\phi}^3+
 \frac{3\tilde{H}}{\tilde{\gamma}^2}\tilde{\phi}^\prime+
 \left(\tilde{\phi}-\frac{4}{\tilde{\lambda}}\tilde{\phi}^3
 \right)\frac{1}{\tilde{\gamma}^3}=0\,.
 \ee


 \begin{center}
 \begin{figure}[htbp]
 \centering
 \includegraphics[width=0.475\textwidth]{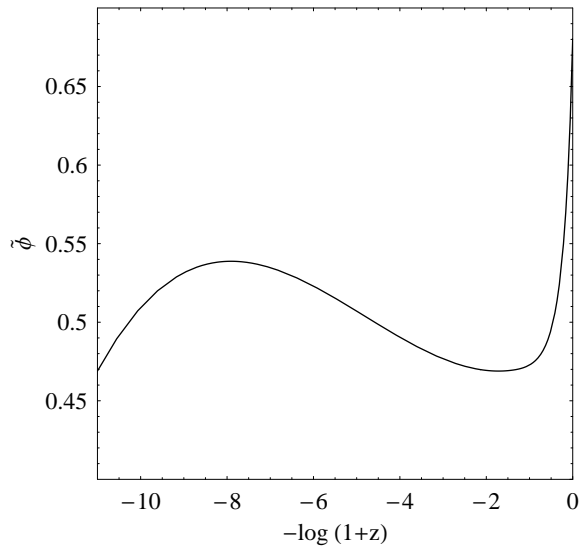}
 \caption{\label{fig1} The solution $\tilde{\phi}$ versus
 $\log a$, for the case of $\tilde{\gamma}_0=\gamma_0=10$.
 See text for details.}
 \end{figure}
 \end{center}


We can find out $\tilde{\phi}(\tilde{t})$ and $a(\tilde{t})$
 by numerically solving the coupled differential equations
 (\ref{eq16}) and (\ref{eq19}), with Eqs.~(\ref{eq17}) and
 (\ref{eq18}). For convenience, as in~\cite{r4,r34}, we adopt
 $\tilde{m}=1$, $\tilde{\lambda}=1$, and $g_{_{\rm YM}}=1$. The
 initial conditions are chosen to be $\Omega_{\phi 0}=0.72$,
 $\Omega_{m0}=0.27$ and hence $\Omega_{r0}=0.01$. As is well
 known, they are fully consistent with cosmological
 observations. Of course, the initial condition for $a$ is
 $a_0=1$. The initial conditions for $\tilde{\phi}$ are
 determined by $\Omega_{\phi 0}$, through
 \be{eq20}
 \Omega_{\phi 0}\equiv\frac{\rho_{\phi 0}}{3m_p^2H_0^2}=
 \frac{\tilde{\phi}_0^2}{3}\left[\left(\tilde{\gamma}_0-1
 \right)\tilde{\phi}_0^2+\frac{1}{2}\right].
 \ee
 As mentioned above, DBI scalar field reduces to quintessence
 when $\gamma\to 1+K/T\simeq 1$. To be distinguished from
 quintessence, as an example, here we adopt
 $\tilde{\gamma}_0=\gamma_0=10$ for DBI scalar field. Thus, we
 can find $\tilde{\phi}_0$ from Eq.~(\ref{eq20}). Then, we can
 obtain $\tilde{\phi}^\prime_0$ from Eq.~(\ref{eq18}). Now, we
 can obtain $\tilde{\phi}(\tilde{t})$ and $a(\tilde{t})$ by
 numerically solving the differential equations (\ref{eq16})
 and (\ref{eq19}) with the initial conditions given above. Once
 $\tilde{\phi}(\tilde{t})$ and $a(\tilde{t})$ are ready, it is
 easy to get $\tilde{\phi}(\log a)$, where $\log$ indicates the
 logarithm to base $10$. In Fig.~\ref{fig1}, we present the
 solution $\tilde{\phi}$ versus $\log a$, for the case of
 $\tilde{\gamma}_0=\gamma_0=10$.


 \begin{center}
 \begin{figure}[htbp]
 \centering
 \includegraphics[width=1.0\textwidth]{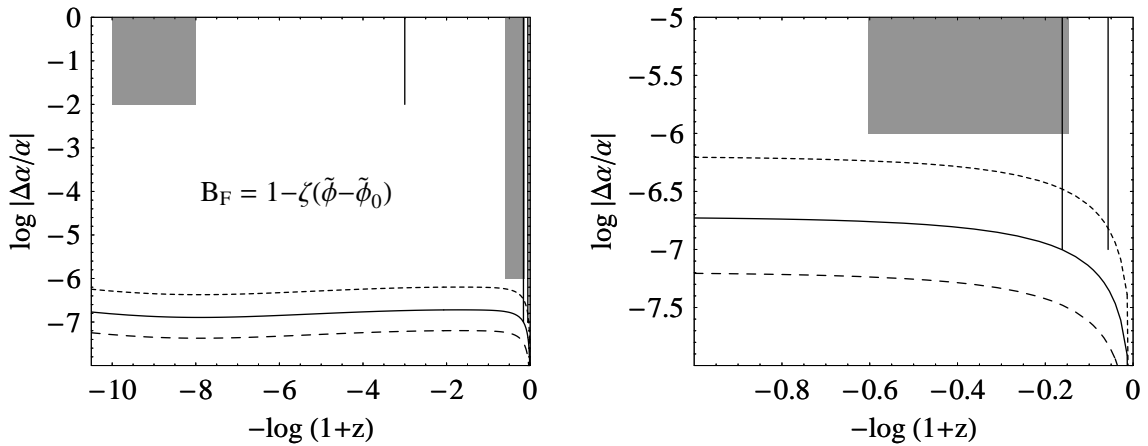}
 \caption{\label{fig2} We plot $\log |\Delta\alpha/\alpha|$ as
 a function of $\log a$ for Case (I) with $\zeta=0.9\times 10^{-6}$
 (solid line), $\zeta=3\times 10^{-6}$ (short-dashed line) and
 $\zeta=0.3\times 10^{-6}$ (long-dashed line). Right panel is
 the enlarged part of $\log a\ge -1$. Only the curves not
 overlapping the gray areas are phenomenologically viable.
 Here, we adopt $\tilde{\gamma}_0=\gamma_0=10$.}
 \end{figure}
 \end{center}


 \vspace{-12mm}  


\subsection{Cosmic evolution of $\alpha$ driven by the DBI
 scalar field with $\tilde{\gamma}_0=\gamma_0=10$}\label{sec3b}

Once the cosmic evolution of DBI scalar field $\tilde{\phi}$
 is on hand, we can easily figure out the corresponding cosmic
 evolution of $\alpha$ from Eqs.~(\ref{eq3}) and (\ref{eq4}),
 if the coupling $B_F(\tilde{\phi})$ is given. In the
 literature, most authors restricted themselves to the case of
 linear coupling for simplicity. Instead, to be general, here
 we consider various coupling $B_F$ following~\cite{r4}. Also,
 we confront the varying $\alpha$ driven by the DBI scalar
 field with the observational constraints mentioned in
 Sec.~\ref{sec1}. In this subsection, we fix
 $\tilde{\gamma}_0=\gamma_0=10$.

 \begin{itemize}

 \item Case (I) Linear coupling\\
 In this case, the coupling is given by
 \be{eq21}
 B_F(\tilde{\phi})=1-\zeta\left(\tilde{\phi}-\tilde{\phi}_0\right),
 \ee
 where $\zeta$ is a constant. This is the mostly considered
 coupling in the literature. From Eq.~(\ref{eq4}), we obtain
 the resulting $\Delta\alpha/\alpha$, and present it in
 Fig.~\ref{fig2}. We tried various $\zeta$ to verify in
 which cases all the observational constraints mentioned
 in Sec.~\ref{sec1} could be simultaneously satisfied. We
 found that they can be all respected for
 $\zeta\leq 0.9\times 10^{-6}$. Notice that in the varying
 $\alpha$ model driven by quintessence~\cite{r4} the upper
 bound of $\zeta$ is $0.6\times 10^{-6}$ for the same $B_F$.
 So, we see that the constraint on $\zeta$ has been relaxed,
 thanks to the relativistic correction of the DBI scalar field.


 \begin{center}
 \begin{figure}[htbp]
 \centering
 \includegraphics[width=1.0\textwidth]{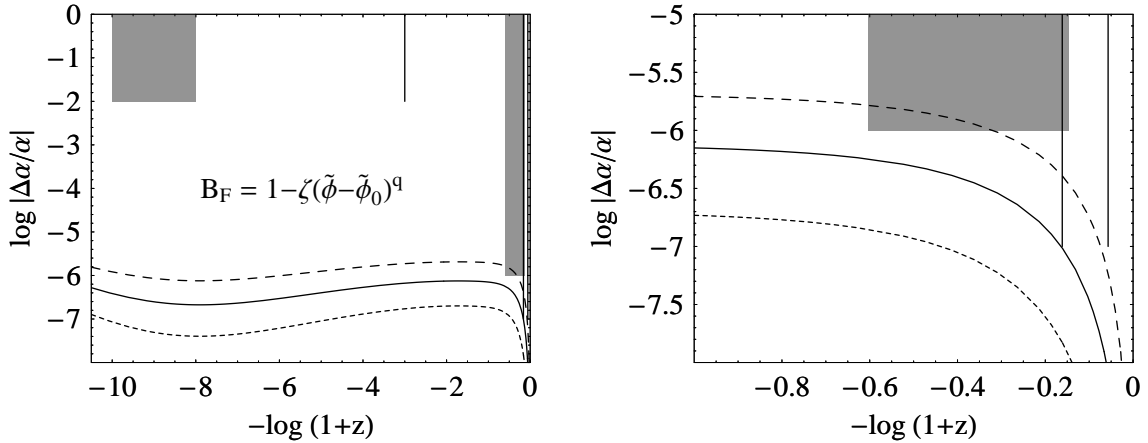}
 \caption{\label{fig3} We plot $\log |\Delta\alpha/\alpha|$ as
 a function of $\log a$ for Case (II) with $\zeta=10^{-4}$ and
 $q=3.15$ (solid line), $q=4$ (short-dashed line) and
 $q=2.5$ (long-dashed line). Right panel is
 the enlarged part of $\log a\ge -1$. Only the curves not
 overlapping the gray areas are phenomenologically viable.
 Here, we adopt $\tilde{\gamma}_0=\gamma_0=10$.}
 \end{figure}
 \end{center}


 \vspace{-10mm}  

 \item Case (II) Polynomial coupling\\
 In this case, one can generalize Eq.~(\ref{eq21}) to
 \be{eq22}
 B_F(\tilde{\phi})=1-\zeta\left(\tilde{\phi}-
 \tilde{\phi}_0\right)^q,
 \ee
 which allows the exponent $q$ to be free. In this case, we
 find that the observations cannot put any upper bound on the
 exponent $q$. In Fig.~\ref{fig3}, we present the resulting
 $\Delta\alpha/\alpha$ for a fixed $\zeta=10^{-4}$ and various
 $q$. We find that the observational constraints can be all
 respected for $q\geq 3.15$. Notice that in the varying
 $\alpha$ model driven by quintessence~\cite{r4} the lower
 bound of $q$ is $6$ for the same $B_F$ with the same
 $\zeta=10^{-4}$. So, we see that the constraint on $q$ has
 been relaxed, thanks to the relativistic correction of the
 DBI scalar field. On the other hand, we find that the upper
 bound of $\zeta$ can be relaxed by increasing $q$. In fact,
 the fine tuning in $\zeta$ can be reduced for the enough
 large $q$. For example, as shown in Fig.~\ref{fig4}, with
 $q=8.4$ the observational constraints can be all respected
 even for $\zeta=1$. Notice that in the varying $\alpha$
 model driven by quintessence~\cite{r4} the lower bound of
 $q$ is $17$ for the same $B_F$ with the same $\zeta=1$.
 Again, the constraint on $q$ has been relaxed in the case
 of DBI scalar field.

 \item Case (III) Power-law coupling\\
 In this case, the coupling under consideration reads
 \be{eq23}
 B_F(\tilde{\phi})=\left(\frac{\tilde{\phi}}{\tilde{\phi}_0}
 \right)^\epsilon,
 \ee
 where $\epsilon$ is a constant. In Fig.~\ref{fig5}, we present
 the resulting $\Delta\alpha/\alpha$ for various $\epsilon$. We
 find that all the observational constraints can be respected
 for $\epsilon\leq 5.5\times 10^{-7}$. Notice that in the
 varying $\alpha$ model driven by quintessence~\cite{r4} the
 upper bound of $\epsilon$ is $4\times 10^{-7}$ for the same
 $B_F$. Again, we see that the constraint on $\epsilon$ has
 been relaxed, thanks to the relativistic correction of the DBI
 scalar field.

 \item Case (IV) Exponential coupling\\
 In this case, the coupling is given by
 \be{eq24}
 B_F(\tilde{\phi})=e^{-\zeta\left(\tilde{\phi}-
 \tilde{\phi}_0\right)}.
 \ee
 Notice that $\tilde{\phi}-\tilde{\phi}_0$ is of order unity
 (see Fig.~\ref{fig1}), if $\zeta$ is of order unity or even
 larger, $B_F$ deviates from $1$ considerably, and it is
 impossible to satisfy all the observational constraints
 at the same time [cf. Eqs.~(\ref{eq3}) and (\ref{eq4})]. If
 $\zeta\ll 1$, we see that $B_F(\tilde{\phi})=e^{-\zeta\left(
 \tilde{\phi}-\tilde{\phi}_0\right)}\simeq 1-\zeta\left(
 \tilde{\phi}-\tilde{\phi}_0\right)$, and hence Case~(IV)
 reduces to Case~(I) considered above.

 \end{itemize}


 \begin{center}
 \begin{figure}[htbp]
 \centering
 \includegraphics[width=1.0\textwidth]{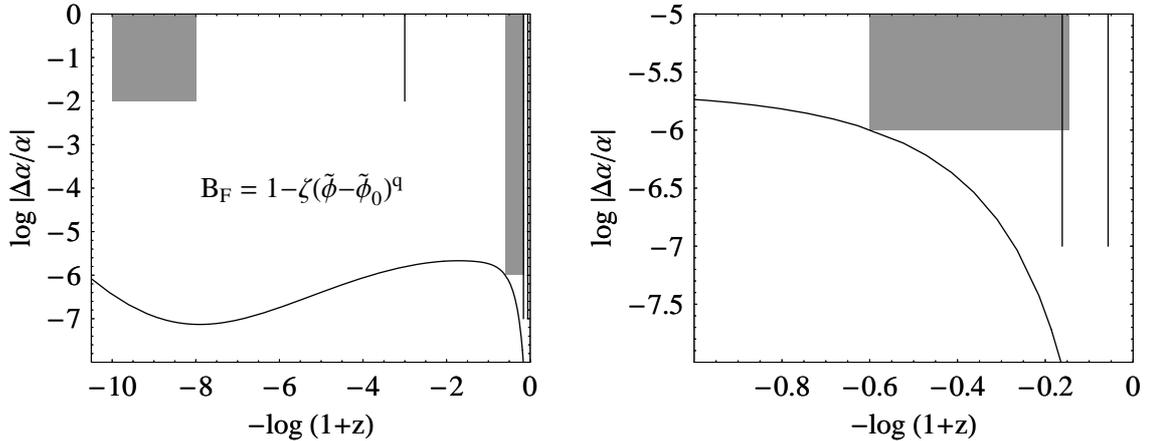}
 \caption{\label{fig4} We plot $\log |\Delta\alpha/\alpha|$
 as a function of $\log a$ for Case (II) with $\zeta=1$ and
 $q=8.4$ (solid line). Right panel is the enlarged part of
 $\log a\ge -1$. Only the curves not overlapping the gray
 areas are phenomenologically viable. Here, we adopt
 $\tilde{\gamma}_0=\gamma_0=10$.}
 \end{figure}
 \end{center}


\vspace{-10mm}  


 \begin{center}
 \begin{figure}[htbp]
 \centering
 \includegraphics[width=1.0\textwidth]{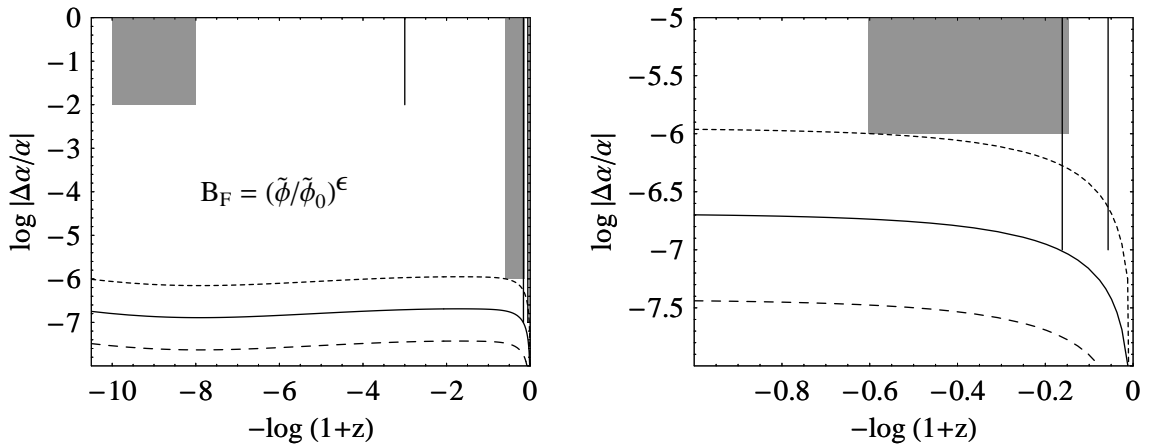}
 \caption{\label{fig5} We plot $\log |\Delta\alpha/\alpha|$ as
 a function of $\log a$ for Case (III) with
 $\epsilon=5.5\times 10^{-7}$ (solid line),
 $\epsilon=3\times 10^{-6}$ (short-dashed line) and
 $\epsilon=1\times 10^{-7}$ (long-dashed line). Right panel is
 the enlarged part of $\log a\ge -1$. Only the curves not
 overlapping the gray areas are phenomenologically viable.
 Here, we adopt $\tilde{\gamma}_0=\gamma_0=10$.}
 \end{figure}
 \end{center}


\vspace{-12mm}  


\subsection{Cosmic evolution of $\alpha$ driven by the DBI
 scalar field with various $\tilde{\gamma}_0$}\label{sec3c}

In the previous subsection, we considered the cosmic evolution
 of $\alpha$ driven by the DBI scalar field with a fixed
 $\tilde{\gamma}_0=\gamma_0=10$. In this subsection, to see the
 relevance of the relativistic correction of the DBI scalar
 field (which is measured by the Lorentz factor $\gamma$), we
 consider the cases with various $\tilde{\gamma}_0$.

\begin{center}
 \begin{figure}[htbp]
 \centering
 \includegraphics[width=0.475\textwidth]{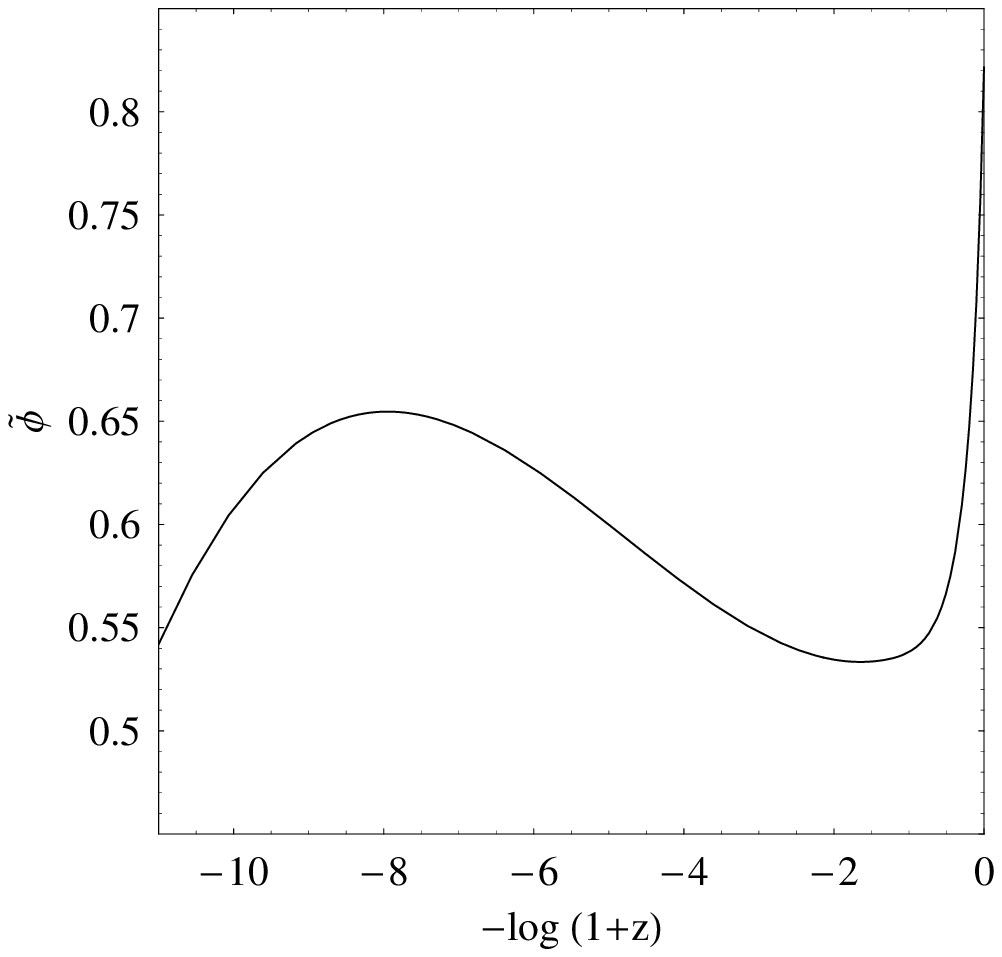}\hfill
 \includegraphics[width=0.475\textwidth]{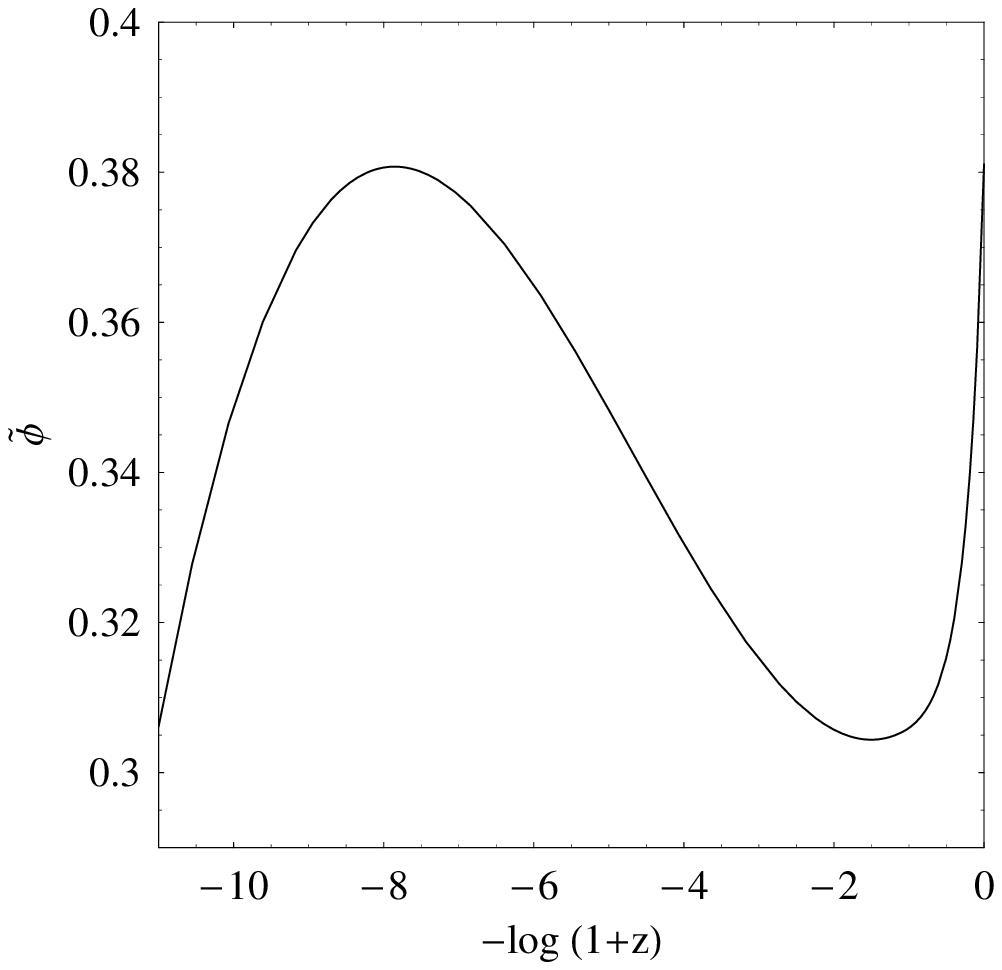}
 \caption{\label{fig6} The solution $\tilde{\phi}$ versus
 $\log a$, for the case of $\tilde{\gamma}_0=\gamma_0=5$
 (left panel) and 100 (right panel).}
 \end{figure}
 \end{center}

\begin{center}
 \begin{figure}[htbp]
 \centering
 \includegraphics[width=1.0\textwidth]{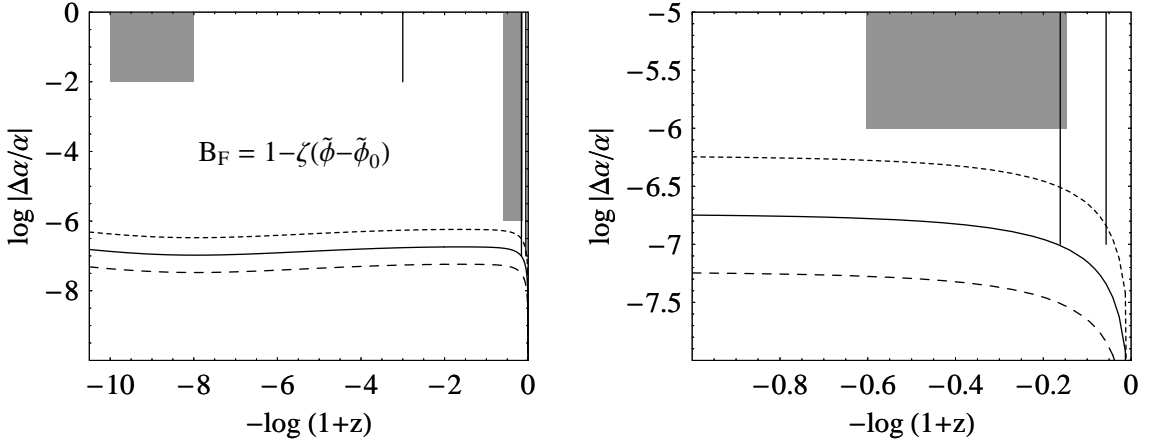}
 \caption{\label{fig7} We plot $\log |\Delta\alpha/\alpha|$ as
 a function of $\log a$ for Case (I) with $\zeta=0.63\times 10^{-6}$
 (solid line), $\zeta=2\times 10^{-6}$ (short-dashed line) and
 $\zeta=0.2\times 10^{-6}$ (long-dashed line). Right panel is
 the enlarged part of $\log a\ge -1$. Only the curves not
 overlapping the gray areas are phenomenologically viable.
 Here, we adopt $\tilde{\gamma}_0=\gamma_0=5$.}
 \end{figure}
 \end{center}

For simplicity, we only consider the linear coupling $B_F$
 given in Eq.~(\ref{eq21}). At first, we adopt
 $\tilde{\gamma}_0=\gamma_0=5$. Following the procedure
 described in the end of Sec.~\ref{sec3a}, we can obtain the
 numerical solution $\tilde{\phi}$ versus $\log a$, and present
 it in the left panel of Fig.~\ref{fig6}. Then, from
 Eq.~(\ref{eq4}), we get the resulting $\Delta\alpha/\alpha$,
 and present it in Fig.~\ref{fig7}. We tried various $\zeta$
 to verify in which cases all the observational constraints
 mentioned in Sec.~\ref{sec1} could be simultaneously
 satisfied. We found that they can be all respected for
 $\zeta\leq 0.63\times 10^{-6}$. Although the upper bound of
 $\zeta$ is still larger than the upper bound
 $0.6\times 10^{-6}$ in the varying $\alpha$ model driven by
 quintessence~\cite{r4}, they are fairly close in fact. This
 is not surprising. As mentioned in Sec.~\ref{sec3a}, the
 DBI scalar field becomes closer to quintessence when $\gamma$
 is smaller~\cite{r28,r33,r34}.


\vspace{-4mm}  

\begin{center}
 \begin{figure}[htbp]
 \centering
 \includegraphics[width=1.0\textwidth]{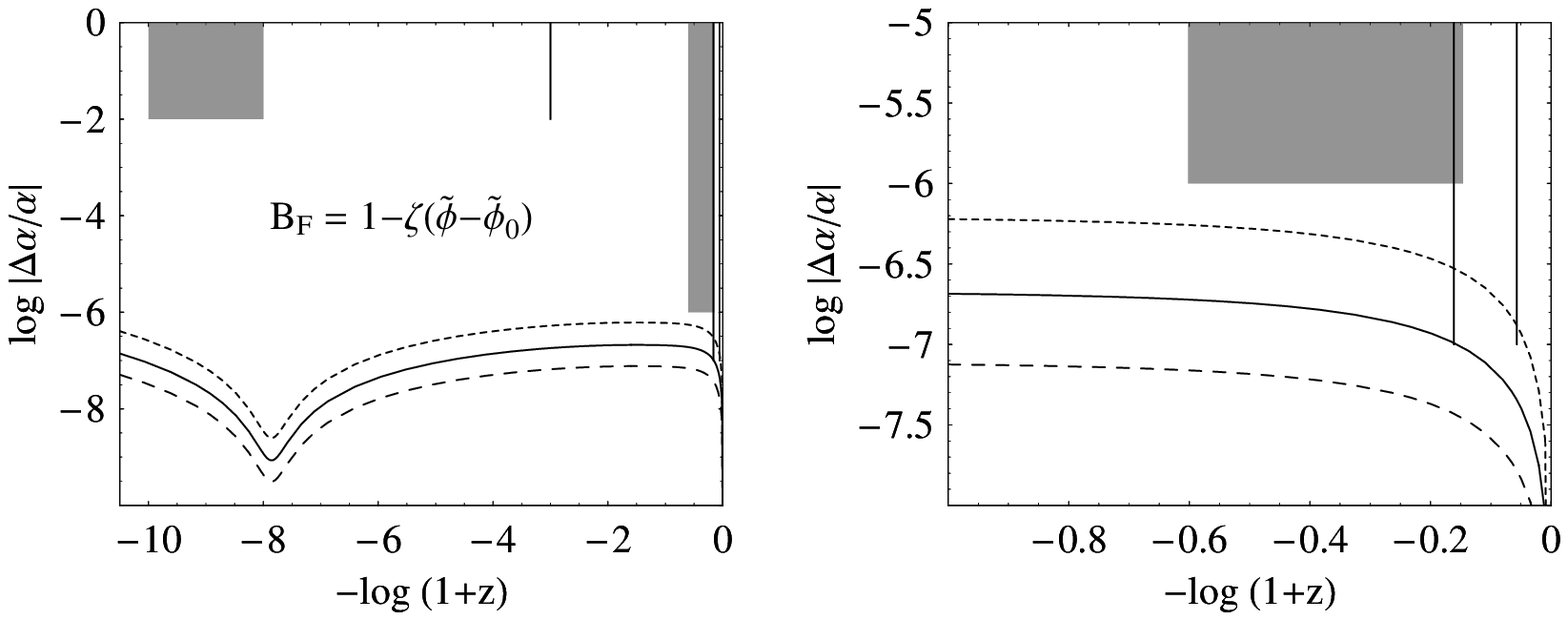}
 \caption{\label{fig8} We plot $\log |\Delta\alpha/\alpha|$ as
 a function of $\log a$ for Case (I) with $\zeta=2.75\times 10^{-6}$
 (solid line), $\zeta=8\times 10^{-6}$ (short-dashed line) and
 $\zeta=1\times 10^{-6}$ (long-dashed line). Right panel is
 the enlarged part of $\log a\ge -1$. Only the curves not
 overlapping the gray areas are phenomenologically viable.
 Here, we adopt $\tilde{\gamma}_0=\gamma_0=100$.}
 \end{figure}
 \end{center}

Other other hand, we consider the case of
 $\tilde{\gamma}_0=\gamma_0=100$. We plot the corresponding
 numerical solution $\tilde{\phi}$ versus $\log a$ in the
 right panel of Fig.~\ref{fig6}, and also present the resulting
 $\Delta\alpha/\alpha$ in Fig.~\ref{fig8}. We find that all
 the observational constraints can be respected
 for $\zeta\leq 2.75\times 10^{-6}$. Obviously, the constraint
 on $\zeta$ has been significantly relaxed, comparing with both
 the cases of DBI scalar field with
 $\tilde{\gamma}_0=\gamma_0=10$ and quintessence~\cite{r4}.
 This is due to the ultra-relativistic effect which is measured
 by the large $\gamma_0=100$. The DBI scalar field
 significantly deviates from quintessence when $\gamma$ is
 fairly large~\cite{r28,r33,r34}.

Together with the results of $\tilde{\gamma}_0=\gamma_0=5$,
 $10$ and $100$, we can clearly see that the relaxation of the
 constraints on the parameters of coupling is mainly due to the
 relativistic correction of the DBI scalar field, which is
 measured by the Lorentz factor $\gamma$. The DBI
 scalar field deviates from quintessence more significantly
 when $\gamma$ is larger~\cite{r28,r33,r34}; and hence as we
 have shown above, the constraints on the parameters of
 coupling is looser.

\vspace{-2mm}  


\section{Conclusion}\label{sec4}
Since about ten years ago, varying $\alpha$ theories attracted
 many attentions, mainly due to the first observational
 evidence from the quasar absorption spectra that the fine
 structure ``constant'' might change with cosmological
 time~\cite{r2,r3}. In this work, we investigated the cosmic
 evolution of $\alpha$ driven by the DBI scalar field. To be
 general, we considered various couplings between the DBI
 scalar field and the electromagnetic field. We also confronted
 the resulting $\Delta\alpha/\alpha$ with the observational
 constraints, and found that various cosmological evolution
 histories of $\Delta\alpha/\alpha$ are allowed. Comparing with
 the case of varying $\alpha$ driven by quintessence~\cite{r4},
 the corresponding constraints on the parameters of coupling
 have been relaxed, thanks to the relativistic correction of
 the DBI scalar field.


\section*{ACKNOWLEDGEMENTS}
We thank the anonymous referee for quite useful comments and
 suggestions, which helped us to improve this work. We are
 grateful to Professors Rong-Gen~Cai and Shuang-Nan~Zhang
 for helpful discussions. We also thank Minzi~Feng, as well as
 Mingzhe~Li, Xiulian~Wang, Zong-Kuan~Guo, and Xin~Zhang, for
 kind help and discussions. This work was supported by the
 Excellent Young Scholars Research Fund of Beijing Institute
 of Technology.

\vspace{-1mm}  


\vspace{-10.5mm}  


\vspace{-10.5mm}  

\renewcommand{\baselinestretch}{1.4}


\end{document}